\documentclass[conference]{IEEEtran}
\IEEEoverridecommandlockouts
\usepackage{cite}
\usepackage{amsmath,amssymb,amsfonts}
\usepackage{algorithmic}
\usepackage{graphicx}
\graphicspath{ {figures/} }
\usepackage{textcomp}

\usepackage{listings}
\usepackage{array}
\usepackage{color}
\usepackage{soul}

\newcolumntype{C}[1]{m{#1}}

\def\BibTeX{{\rm B\kern-.05em{\sc i\kern-.025em b}\kern-.08em
    T\kern-.1667em\lower.7ex\hbox{E}\kern-.125emX}}
\begin{document}

\bstctlcite{IEEEexample:BSTcontrol}

\title{The Effectiveness of Low-Level Structure-based Approach Toward Source Code Plagiarism Level Taxonomy
}

\author{\IEEEauthorblockN{Oscar Karnalim}
\IEEEauthorblockA{Faculty of Information Technology \\
Maranatha Christian University\\
Bandung, Indonesia \\
oscar.karnalim@it.maranatha.edu}
\and
\IEEEauthorblockN{Setia Budi}
\IEEEauthorblockA{Faculty of Information Technology \\
Maranatha Christian University\\
Bandung, Indonesia \\
setia.budi@it.maranatha.edu}
}

\maketitle

\begin{abstract}
Low-level approach is a novel way to detect source code plagiarism. 
Such approach is proven to be effective when compared to baseline approach (i.e., an approach which relies on source code token subsequence matching) in controlled environment. 
We evaluate the effectiveness of state of the art in low-level approach based on Faidhi \& Robinson's plagiarism level taxonomy; 
real plagiarism cases are employed as dataset in this work. 
Our evaluation shows that state of the art in low-level approach is effective to handle most plagiarism attacks. 
Further, it also outperforms its predecessor and baseline approach in most plagiarism levels.
\newline
\end{abstract}

\section{INTRODUCTION}
Source code plagiarism is an act of reusing others' code without acknowledging the original author(s) \cite{Cosma2008}.
It is an emerging issue among undergraduate students in Computer Science (CS); 
since most assignments in CS are related to programming and they are relatively easy to be replicated \cite{Karnalim2017IAENG}. 
In addition, source code plagiarism is difficult to be detected and the cases are not limited among students with poor academic performance only \cite{Rabbani2017}. 
In response to this issue, a number of plagiarism detection systems have been proposed \cite{Lancaster2004}.
These systems are expected to handle numerous source codes and detect complex plagiarism cases efficiently with accurate result. 

One recent work in source code plagiarism is by adopting low-level structure-based approach.
This technique relies on low-level representation to measure similarity in source code \cite{Karnalim2016}.
Such representation could result to a better accuracy compared to baseline approach (i.e., an approach which determines similarity based on source code subsequence matching) since there are no syntactic-sugar forms and delimiter tokens involved \cite{Karnalim2016,Karnalim2017ICSESS,Karnalim2017IAENG}. 
One comprehensive work which implement low-level structure-based approach is presented in \cite{Karnalim2017IAENG}; a wide range of plagiarism aspects in Java source code are covered in the work.

This paper serves as an extension of the work presented in \cite{Karnalim2017IAENG}, by providing an in-depth evaluation toward the proposed approach on real plagiarism cases.
A plagiarism taxonomy proposed in \cite{FaidhiJ.A.W;Robinson1987} is adopted in this work.
Such taxonomy has been widely accepted in the field of source code plagiarism detection \cite{Kustanto2009,Prechelt2002,Ganguly2017,Novak2016,Karnalim2017THESTE}.
\section{RELATED WORKS}
Source code plagiarism detection techniques can be classified into three categories: attribute-based, structure-based, and hybrid approach \cite{Al-Khanjari2010,Karnalim2017IAENG}. 
Attribute-based approach measures similarity based on key properties extracted from source codes (e.g., the number of identifier and line of code).
Structure-based approach quantifies similarity based on the structure of the code (e.g., token subsequence). 
Hybrid approach combines the former two categories.

This paper specifically focuses on low-level structure-based approach where similarity in source code is measured based on the structure of low-level tokens (i.e., tokens extracted from compiled form of given source code). 
It has been adopted in \cite{juri2011a,jurivcic2011performance,Rabbani2017,Ji2008,Karnalim2017IAENG,Karnalim2016,Karnalim2017ICSESS} and designed to handle either .NET or Java programming language.
The works on .NET programming language rely on Common Intermediate Language (i.e., .NET's low-level representation) where several different techniques to measure similarity in the tokens were adopted. 
Levenstein distance was applied in \cite{juri2011a}; whereas Running-Karp-Rabin Greedy-String-Tiling algorithm and adaptive local alignment were adopted in \cite{jurivcic2011performance} and \cite{Rabbani2017} respectively. 
In contrast, the works on Java programming language rely on bytecodes (i.e., Java's low-level representation) by considering various programming features \cite{Ji2008,Karnalim2016,Karnalim2017ICSESS,Karnalim2017IAENG}.
A work proposed in \cite{Ji2008}, at some extent, becomes a baseline for other works on Java programming language. 

A work proposed in \cite{Karnalim2016} is extended from \cite{Ji2008} by incorporating four additional features: instruction generalization, instruction reinterpretation, method-based comparison, and modified method linearization. 
The first two features omit over-technical detail in the bytecode token sequence by replacing several tokens with a more-simplified form (e.g., \textit{switch-case} token sequence is replaced with a standard \textit{goto}-based sequence).
The last two features reduce the number of false-positive tokens by considering token context. 
Instead of comparing the whole token sequence at once, it compares the sequences locally per method pair wherein each method invocation is linearized according to the content of the invoked method. 

Considering the benefits of \cite{Karnalim2016}, works proposed in \cite{Karnalim2017ICSESS} and \cite{Karnalim2017IAENG} extend that work.
The former work incorporates a naive solution for abstract method linearization; 
it considers the content of all candidate methods as a replacement of an abstract method invocation.
The latter work incorporates three additional features: flow-based token weighting, argument removal heuristic, and invoked method removal. 
The first two features contribute to the effectiveness of the proposed approach by generating more accurate similarity result. 
Flow-based token weighting is used to differentiate similar tokens with different scope while argument removal heuristic is used to remove remaining arguments at method linearization phase. 
In addition, invoked method removal improves the efficiency (i.e., reduces processing time) by excluding the content of invoked methods from comparison.

Works proposed in \cite{Karnalim2016}, \cite{Karnalim2017ICSESS}, and \cite{Karnalim2017IAENG} use a source-code-token approach as comparison baseline. 
Such approach works in threefold: converts both source codes into lexical token sequences, removes the comments, and compares resulted token sequences using maximum matching similarity \cite{Prechelt2002} with Running-Karp-Rabin Greedy-String-Tiling algorithm \cite{Wise1995}. 
According to these works, low-level approach outperforms the baseline approach in terms of effectiveness.
Compilation phase (which is exclusively conducted by low-level approach to translate source code to low-level tokens) generates three benefits:
\begin{enumerate}
\item Resilient to comment, whitespace, and delimiter modification; tokens related to these modifications are excluded. 
\item Resilient to local variable renaming and syntactic-based modification; local variables are automatically renamed and most syntactic-sugar forms will be translated to their original form.
\item Generate less mismatched tokens; in terms of instruction representation, low-level token sequence is more concise compared to source code token sequence.
\end{enumerate}

In the field of source code plagiarism detection, plagiarism taxonomy \cite{FaidhiJ.A.W;Robinson1987} is commonly used as an evaluation metric.
Such metric has been implemented in number of studies  \cite{Karnalim2016,Rabbani2017,Kustanto2009,Prechelt2002,Ganguly2017,Novak2016,Karnalim2017THESTE}.
Difficulty level with a range from level 1 to level 6 is applied in the metric to represent a spectrum of difficulty from the easiest to the hardest one; 
where signature attacks from each level are inclusive toward higher level categories.
Table \ref{tab:plag_level} presents the attack signature for each level including its example.

\begin{table*}[htbp]
\caption{Plagiarism Levels defined in \cite{FaidhiJ.A.W;Robinson1987}}
\label{tab:plag_level}
\centering
\begin{tabular}{| C{0.05\textwidth}| C{0.45\textwidth}| C{0.4\textwidth}|}
\hline
\bfseries Level & \bfseries Attack Signatures & \bfseries Example\\
\hline
1
& 
Comment and whitespace modification 
&
Removing all comments from given source code \\
\hline
2
& 
Identifier modification (i.e., changing lexical name from one to another)
&
Renaming all local variables \\
\hline
3
& 
Component declaration relocation
&
Moving all variable declarations to the beginning of main method \\
\hline
4
& 
Method structure change
&
Replacing all method invocations with their respective invoked-method's content \\
\hline
5
& 
Program statement replacement (i.e., changing statements with other statements that share similar semantic yet different syntactic form)
&
Replacing \textit{while} statement with \textit{for} statement\\
\hline
6
& 
Logic change (i.e., changing statements with other statements that share no similarity in terms of syntactic and semantic form)
&
Replacing an iterative traversal with the recursive one that generates similar result\\
\hline
\end{tabular}
\end{table*}

\section{METHODOLOGY}
The implementation of Faidhi \& Robinson’'s taxonomy \cite{FaidhiJ.A.W;Robinson1987} for evaluation in most low-level source code plagiarism studies \cite{Rabbani2017,Karnalim2016,Karnalim2017IAENG} are limited to a controlled environment, where each plagiarism case contains only a single plagiarism attack. 
As a consequence, those studies may suffer from lack of real life application, considering such controlled dataset excludes combined attacks (which are commonly found in real life). 
In contrast to those works, we utilize real plagiarism cases (captured from undergraduate students without limiting involved plagiarism attacks) as a dataset in our exploration.
Such dataset would enable us to analyze the impact of low-level approach toward combined attacks and unexpected cases.
Our dataset is filtered from raw data proposed in \cite{Karnalim2016} (where plagiarism cases have been mapped into Faidhi \& Robinson’'s taxonomy \cite{FaidhiJ.A.W;Robinson1987}, based on the highest plagiarism attack level included) by manually removing misclassified cases.
It consists of 355 plagiarism cases with each plagiarism level covers between 56 and 63 cases.

Three different approaches are considered in our evaluation: Extended Low-Level Approach (Ext-LLA), Low-Level Approach (LLA), and Source-code-Token Approach (STA). 
Ext-LLA \cite{Karnalim2017IAENG} is a state of the art in low-level structure-based approach which impact will be measured in this study.
LLA \cite{Karnalim2016} is the predecessor of Ext-LLA; 
its result will be compared to Ext-LLA's for measuring the impact of Ext-LLA's signature features.
STA is a comparison baseline approach used in \cite{Karnalim2016,Karnalim2017ICSESS,Karnalim2017IAENG}; 
its result will be compared to Ext-LLA's for measuring the impact of low-level representation.

\begin{figure}[b]
\centerline{\includegraphics[width=0.45\textwidth]{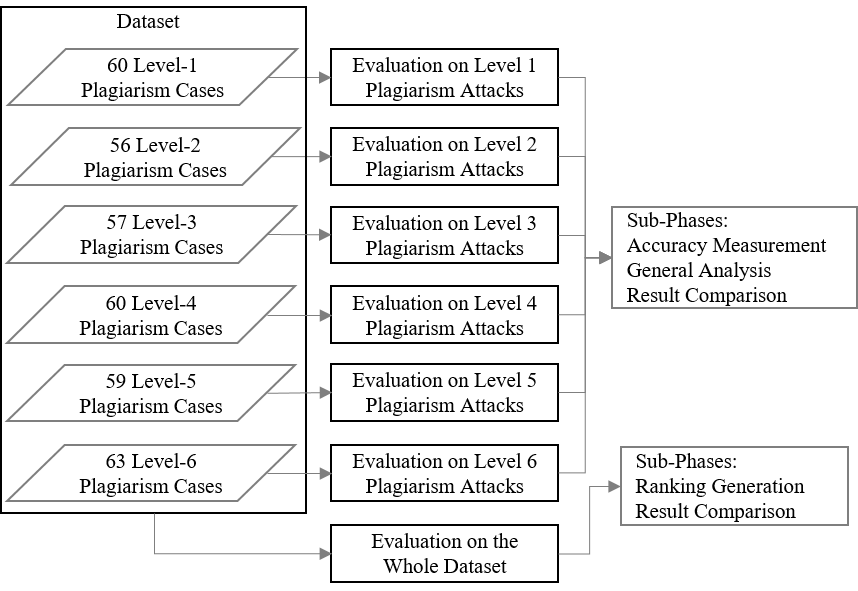}}
\caption{Evaluation methodology. The first six phases evaluate Ext-LLA's effectiveness locally per plagiarism level while the last phase evaluates its effectiveness in general.}
\label{fig:method_overview}
\end{figure}

In general, our methodology consists of seven evaluation phases (as illustrated in Fig. \ref{fig:method_overview}). 
The first six phases incorporate three sub-phases: accuracy measurement, general analysis, and result comparison. 
Accuracy measurement is conducted by generating Reversed number of Mismatched Token (RMT) for each approach per case from our dataset. 
It is calculated by negating the number of Mismatched Token (MT) from both token sequences (A and B), resulting a non-positive integer which is ranged from $-\infty$ to 0 (as formulated in (\ref{eq:RMT})).  

\begin{equation}
RMT(A,B) = -1 * MT(A,B)
\label{eq:RMT}
\end{equation}

It is important to note that RMT is preferred as our effectiveness metric instead of normalized similarity (which is commonly used in the works of source code plagiarism detection \cite{Prechelt2002,Kustanto2009}) since RMT is not affected by the number of involved token.
The number of involved token may obfuscate the result considering source code always has more tokens when compared to low-level code, even though both codes refer to similar semantic \cite{Karnalim2016,Rabbani2017}.
In fact, RMT has been used in previous works about low-level approach \cite{Karnalim2016,Karnalim2017ICSESS,Karnalim2017IAENG} despite the use of different terminology (we use RMT as our terminology since we would argue it is the most appropriate name that represents how it works).

General analysis is conducted by analyzing the characteristics of Ext-LLA when handling plagiarism attacks on given level. 
Further, result comparison is conducted by comparing the trends between Ext-LLA and other two approaches toward given plagiarism level. 

The $7^{th}$ phase is conducted in two sub-phases: ranking generation and result comparison. 
A rank is assigned for each approach per case in ranking generation sub-phase, where a high rank implies high RMT. 
It is important to note that a particular rank is not exclusively assigned to one approach, considering two or more approaches may yield a same RMT value.
Further, result comparison is conducted by comparing the trends between involved approaches from ranking perspective. 
Both phases will be conducted for the whole dataset (i.e., a merged form of six level-based dataset as illustrated in Fig. \ref{fig:method_overview}).

\section{RESULT AND DISCUSSION}
\subsection{Evaluation Toward Level-1 Plagiarism Attacks}
Based on observation toward our dataset, we identified 60 plagiarism cases related to level-1 plagiarism attacks (which is about comment \& whitespace modification). 
Having comment and whitespace tokens excluded at compilation phase, level-1 attacks in our dataset are accurately detected (i.e., zero RMT for all level-1 cases) with Ext-LLA. 
Therefore, we can assure that Ext-LLA is resistible to level-1 plagiarism attacks.

In comparison to STA, Ext-LLA generates higher RMT in most cases even though both approaches exclude comment and whitespace tokens. 
Further investigation shows that STA’'s low result is caused by the existence of IDE- and ownership-related modification. 
IDE-related modification refers to a modification that is automatically generated when the code is imported to other IDE (e.g., automatically-generated package name) whereas ownership-related modification refers to a modification that is required to claim the ownership for given code (e.g., renaming main class name with student ID). 
Both modifications generate a slight difference on source code level, resulting lower RMT for STA.  
However, since both modifications are out of level-1 attack scope, it cannot be stated that Ext-LLA outperforms STA.

In contrast to STA, LLA generates a fairly similar result to Ext-LLA in all cases;
comment and whitespace tokens are removed during compilation phase in both approaches.

\subsection{Evaluation Toward Level-2 Plagiarism Attacks}
From our dataset, we identified 56 plagiarism cases related to level-2 plagiarism attacks.
We also found that identifier modification (a level-2 signature attack) in our dataset occurs either in the form of local variable or method name modification. 
Ext-LLA, at some extent, is able to handle given attacks accurately;
it generates zero RMT for all level-2 cases.
Such performance is achieved due to Bytecode’'s local variable renaming (which renames all local variables with technical names based on their first occurrence) and method linearization (which removes method name as a result of linearization process). 
These mechanisms handles both local variable and method name modification respectively.
Therefore, we can assure that Ext-LLA is also resistible to level-2 plagiarism attacks.

STA compares source code token based on its mnemonic and considers each occurrence of renamed identifier as a mismatch;
it is in contrast to Ext-LLA which does not directly compare source code token mnemonic. 
Pre-processing mechanisms are applied in Ext-LLA to mitigate the number of mismatches in advance.
In our evaluation, we found Ext-LLA generates higher RMT in all cases compared to STA.

Both LLA and Ext-LLA handle level-2 plagiarism in a similar fashion.
Therefore, it is expected that they both yield a similar result in our evaluation study.

\subsection{Evaluation Toward Level-3 Plagiarism Attacks}
There are 57 plagiarism cases related to level-3 plagiarism attacks identified from our dataset.
Level-3 attacks occur in the form of relocation in either variable declaration (i.e., relocating variable declaration within the same scope or to a larger scope) or method declaration (i.e., restructuring method declaration). 
In general, Ext-LLA is resistible to those attacks except on two conditions: relocating variable declaration from local to class scope and relocating variable declaration from a looping body or a branching body to its larger scope. 
The former condition relocates variable declaration from main method to implicit constructor, therefore breaking down these methods'’ matched sequences to shorter sequences that are mostly undetected (since their length is below Ext-LLA's minimum matching length).
The latter condition alters the token weight of a relocated variable. 
Ext-LLA is sensitive to changes in token weight; 
it will assume that two tokens are within two different scopes and they can not be considered matched. 
In our evaluation study, Ext-LLA yields zero RMT for 46 out of 57 cases.

Evaluation toward our level-3 dataset results to higher RMT in Ext-LLA compared to in STA.
However, from our investigation, we discovered that component declaration relocation does not have significant effect on STA. 
Low RMT in STA is more likely affected by identifier modification (one of level-2 signature attacks). 
Such side effect is reasonable considering level-3 attacks also inherently covers plagiarism attacks in lower levels.

While handling level-3 attacks, both LLA and Ext-LLA yield similar result in most cases.
They only perform differently in the case where a variable declaration is relocated from loop body to its outer scope. 
In Ext-LLA, relocated variable declaration is considered as mismatched tokens due to its modified scope. 
Therefore it is expected that, while handling level-3 attacks, Ext-LLA is slightly less effective than LLA.

\subsection{Evaluation Toward Level-4 Plagiarism Attacks}
There are 60 plagiarism cases in our dataset which cover level-4 attacks. 
Such attacks are occurred in the form of replacing method invocation with its respective content or encapsulating program statements as a method. 
The former form may omit some local variable declarations by replacing them with existing variables on the invoker method; the latter form may introduce some local variable declarations on newly-created method to smoothly transfer some values from invoker method. 
In our evaluation, we found that Ext-LLA generates a significantly higher RMT than STA (i.e., in average Ext-LLA generates -3 RMT per case while STA generates -14 RMT).
Apart from the high occurrences of local variable modification, such significant performance is also due to Ext-LLA'’s token representation (which is more compact than STA'’s). 

In contrast to LLA, Ext-LLA is exclusively featured with argument removal heuristic (which removes argument-preparation tokens for each method invocation). 
Such heuristic reduces the number of mismatched tokens considering most method invocations in our dataset are featured with argument-preparation tokens.
It is not surprising that, in our evaluation study, Ext-LLA results to better performance compared to LLA; 
it generates higher RMT in 33 of 60 cases.

\subsection{Evaluation Toward Level-5 Plagiarism Attacks}
From our dataset, we discovered 59 plagiarism cases related to level-5 plagiarism attacks.
We also identified two forms of program statement replacement (i.e., level-5 signature attack) in our dataset: replacement with exactly-similar and approximately-similar semantic. 
In exactly-similar semantic, the statement replacement yields similar behavior to the original one on all possible occasions (e.g., replacing \textit{while} traversal with \textit{for} traversal). 
Meanwhile, in approximately-similar semantic, the statement replacement only yields similar behavior at least on one occasion (e.g., replacing \textit{while} traversal with \textit{do-while} traversal). 

In our evaluation study on level-5 attacks, we found that Ext-LLA yields higher RMT compared to STA.
In average, Ext-LLA generates -4 RMT per case while STA generates -20 RMT.
Replacing and replaced tokens on bytecode level is more uniform to each other when compared to the source code level; 
such uniformness favors Ext-LLA to outperform STA.

We also found that Ext-LLA outperforms LLA in 30 cases.
Having argument removal heuristic on board, Ext-LLA manages to exclude some mismatched tokens from argument-preparation tokens. 
In addition, invoked method removal employed by Ext-LLA also manages to exclude some mismatched tokens from the content of the invoked methods. 
Ext-LLA is only underperformed by LLA in one case where a conversion from \textit{while} to \textit{do-while} traversal is involved.
In contrast to LLA, Ext-LLA considers all tokens from both traversal bodies as mismatched since these traversals generate different control flow path. 

\subsection{Evaluation Toward Level-6 Plagiarism Attacks}
From our dataset, there are 63 cases which comply to level-6 plagiarism attacks.
Logic change (which is level-6 signature attack) is difficult to be detected using Ext-LLA; 
in most occasions, different logics are represented with different bytecodes and token scopes. 
Nevertheless, in our evaluation study, we found that Ext-LLA still generates higher RMT than STA. 
In average, Ext-LLA generates -8 RMT per case while STA generates -25 RMT.

Our evaluation study also shows that Ext-LLA outperforms LLA in 28 cases; thanks to argument removal heuristic and invoked method removal (both mechanisms mitigate the number of mismatched tokens in similar fashion as in level-5 attacks). 
Ext-LLA is underperformed by LLA in six cases where the modification of control flow and the existence of operation in method arguments are involved. 
On the one hand, modification in control flow generates lower RMT on Ext-LLA since more mismatched tokens will be generated as a result of Ext-LLA’'s flow-based token weighting. 
However, we would argue that Ext-LLA’'s result in these cases is more sensible than LLA’'s; tokens with different scope should not be considered as similar to each other. 
On the other hand, the existence of operation in method arguments generates lower RMT on Ext-LLA;
argument removal heuristic applied in Ext-LLA cannot correctly handle operation that is implicitly conducted on method arguments. 
\newline
\newline
\newline

\subsection{Evaluation Toward the Whole Dataset}
\begin{figure}[t]
\centerline{\includegraphics[width=0.5\textwidth]{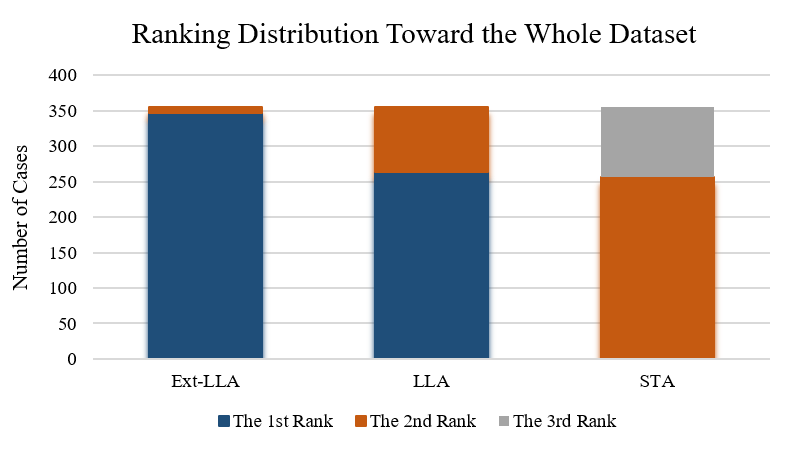}}
\caption{Ranking distribution toward the whole dataset; The color of each bar represents a rank; horizontal axis represents involved approaches; and vertical axis represents the number of cases. 
To assign a rank, given approach will be compared to each other per case in terms of RMT, where higher RMT refers to higher rank; 
if several approaches generate similar RMT, then these approaches will be assigned with the same rank.}
\label{fig:ranking_distribution}
\end{figure}

The ranking distribution of Ext-LLA, LLA, and STA toward the whole dataset can be seen in Fig. \ref{fig:ranking_distribution}. 
Two findings can be highlighted from our evaluation study.
First, Ext-LLA yields the best performance, followed by LLA and STA.
Ext-LLA generates the highest RMT on 347 out of 355 cases. 
Second, three specific features (i.e., flow-based token weighting, argument removal heuristic, and invoked method removal) employed by Ext-LLA are able to enhance the effectiveness of low-level approach;
Ext-LLA generates the highest RMT on more cases than LLA.

\section{CONCLUSION AND FUTURE WORK}
In this paper, a comprehensive evaluation on state of the art in low-level plagiarism detection approach \cite{Karnalim2017IAENG} toward plagiarism level taxonomy (with real plagiarism cases) is presented. 
Five findings can be highlighted from our evaluation study.
First, the approach is resistible to the first two plagiarism attack levels.
Second, RMT resulted from such approach is reversely proportional to increasing plagiarism level on level-3 to level-6 attack category.
Third, the approach outperforms its predecessor and source-code-token approach.
Fourth, the approach is more sensitive to detect false-positive result; it differentiates tokens not only based on their mnemonic but also their scope.
Fifth, signature features proposed in such approach enhance the effectiveness of low-level approach.

For future work, we plan to extend state of the art in low-level approach \cite{Karnalim2017IAENG} for handling source code plagiarism in object-oriented environment. 
Different with works proposed in \cite{Karnalim2017ICSESS,Karnalim2018kingsaud}, we will expand such approach by incorporating attribute-based approach.

\begingroup
\let\itshape\upshape
\bibliographystyle{IEEEtran}
\bibliography{IEEEabrv,references}
\endgroup

\end{document}